\documentclass{article}

\usepackage{spconf,amsmath,graphicx}
\usepackage{dblfloatfix}
\usepackage{placeins}
\usepackage[draft,bookmarks=false]{hyperref}
\usepackage{float}
\usepackage{booktabs}
\usepackage{amssymb}

\usepackage{url}
\title{Covariate-Conditioned One-Step Diffusion for
Multi-Horizon 3D Brain MRI Prediction in
Aging and Neurodegeneration}

\name{Natenaile Asmamaw Shiferaw$^{1}$,
Ali Farki$^{1}$,
Deepika Koundal$^{1,2}$,
Jussi Tohka$^{1}$}

\address{
$^{1}$ A.I. Virtanen Institute for Molecular Sciences,
University of Eastern Finland, Kuopio, Finland\\
$^{2}$ University of Petroleum and Energy Studies (UPES),
Dehradun, India
}
\begin{document}

\maketitle

\begin{abstract}
Predicting longitudinal brain changes from magnetic resonance
imaging (MRI) can help characterize structural progression
associated with aging and neurodegeneration. However,
diffusion-based forecasting typically requires iterative
denoising, limiting inference efficiency. We propose a covariate-conditioned one-step diffusion
framework for multi-horizon prediction of 3D gray-matter
density (GMD) maps. Rather than directly generating future
anatomy, the model predicts longitudinal GMD changes
through residual prediction in a single denoising step.
The predicted changes are added to the baseline GMD map
to reconstruct future anatomy. The model is conditioned
on baseline anatomy, age, diagnosis, APOE4 allele count,
and continuous follow-up time.
Training combines mean squared error with a Pearson
correlation loss to improve the spatial agreement between
predicted and observed GMD changes. A unified model
supports prediction horizons from 2 to 10 years.
On the held-out ADNI test set, our method achieves a
$\Delta$-Pearson of 0.478, mean squared error (MSE) of
$5.176\times10^{-3}$, and structural similarity index
measure (SSIM) of 0.979, with $\Delta$-Pearson reaching
0.582 at 10 years. Without retraining, external evaluation
on AIBL yields a $\Delta$-Pearson of 0.450, MSE of
$4.016\times10^{-3}$, and SSIM of 0.985.
Ablation studies further demonstrate the benefits of
residual prediction, covariate conditioning, and
correlation-based supervision.
Code: \nolinkurl{https://github.com/natenaile/OneStepBrainMRI}.
\end{abstract}

\begin{keywords}
longitudinal brain MRI, diffusion models, one-step inference, multi-horizon prediction
\end{keywords}

\section{Introduction}

Longitudinal brain magnetic resonance imaging (MRI)
characterizes structural changes associated with aging
and neurodegeneration~\cite{lorenzi2015disentangling},
but assessing progression requires follow-up scans over
several years~\cite{tondelli2012structural}. Existing
forecasting methods often predict regional volumes or
biomarkers~\cite{hadji2026predicting}, whereas 3D image
prediction captures the spatial distribution of anatomical
changes~\cite{farki2026forecasting,litrico2024tadm,puglisi2025brain}.
Such prediction remains challenging due to subtle,
heterogeneous changes that depend on individual
characteristics and follow-up time~\cite{hedman2012human}.
Accurate forecasting could support the investigation of
disease progression by providing spatially detailed
estimates of future brain anatomy.

Longitudinal MRI forecasting has been explored using
deterministic models~\cite{farki2026forecasting} and
diffusion approaches with temporal conditioning
~\cite{litrico2024tadm} or latent-space generation
~\cite{puglisi2025brain}. Diffusion models provide a
generative framework for complex longitudinal changes.
However, conventional diffusion requires multiple
denoising steps at inference
~\cite{ho2020denoising,song2020denoising}, increasing
computational cost for 3D volumes.

Recent single-step diffusion methods exploit spatial
correlations between input and target images for
efficient dense prediction while maintaining
diffusion's generative capabilities
~\cite{he2025lotus,ye2024stablenormal,wu2025stableintrinsic}.
The anatomical correspondence between baseline and
follow-up brain MRI scans motivates single-step
longitudinal forecasting. We propose a covariate-conditioned
one-step diffusion framework for multi-horizon 3D
gray-matter density (GMD) prediction.
Our contributions are:

\vspace{-4pt}
\begin{itemize}
    \setlength{\itemsep}{0pt}
    \setlength{\parskip}{0pt}
    \setlength{\parsep}{0pt}

    \item We introduce single-step diffusion for
    longitudinal 3D GMD prediction without iterative
    reverse sampling.

    \item We formulate prediction as residual GMD
estimation conditioned on baseline anatomy, age,
diagnosis, apolipoprotein E $\varepsilon$4 (APOE4)
allele count, and follow-up time
within a unified multi-horizon model.

    \item We evaluate whole-brain and regional prediction
    across multiple follow-up horizons, external
    generalization on an independent cohort, and the
    contributions of individual model components
    through ablation studies.

\end{itemize}
\begin{figure*}[!b]
    \centering
    \includegraphics[width=0.86\textwidth]{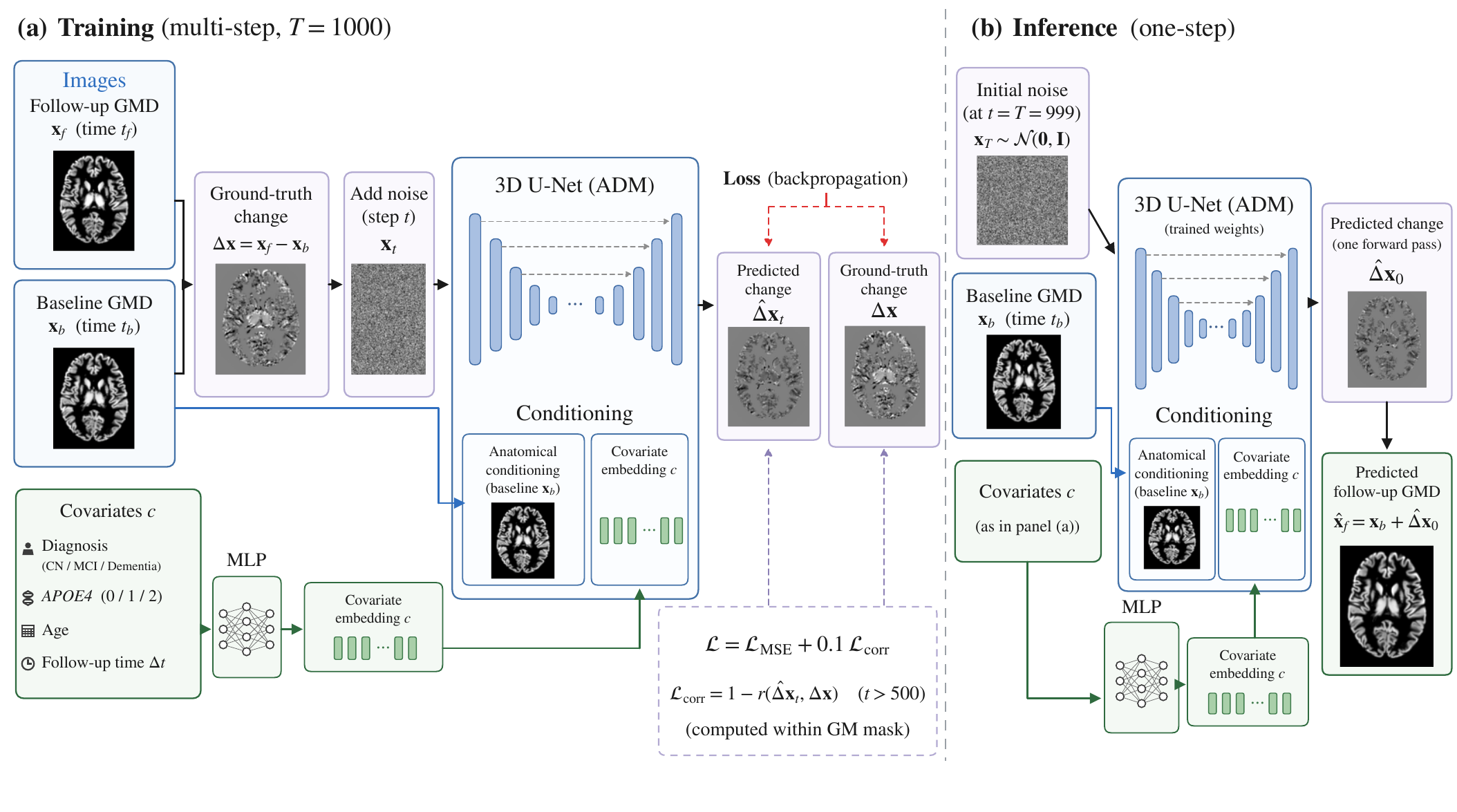}
    \caption{Overview of the proposed framework.
(a) Training with noisy longitudinal GMD changes,
anatomical conditioning, and subject-specific information.
(b) One-step inference for predicting future GMD maps
without iterative denoising.
}
    \label{fig:method}
\end{figure*}
\section{Material}

We use the Alzheimer's Disease Neuroimaging Initiative (ADNI) \cite{adni_data}
for development and internal evaluation, and the Australian
Imaging, Biomarkers and Lifestyle (AIBL) \cite{aibl_data} study for external
validation. 3D T1-weighted MRI scans were processed
using the Computational Anatomy Toolbox (CAT12, v12.8.1)
~\cite{gaser2024cat,cat12website}
to obtain modulated GMD maps, normalized to Montreal
Neurological Institute (MNI) space at $1.5$\,mm isotropic
resolution ($113\times137\times113$ voxels).
ADNI comprised 10,264 scan pairs from 1,468 participants
across 2--10 years, while AIBL included 332 pairs from
150 participants at 3 and 6 years. Diagnostic distributions
(cognitively normal [CN]/mild cognitive impairment
[MCI]/dementia) were 40/49/11\% in ADNI and 79/15/6\%
in AIBL. Further cohort characteristics and
interval-specific sample sizes are given in
Table~\ref{tab:dataset}.

\begin{table}[t]
\centering
\caption{Longitudinal cohort characteristics of ADNI and
the external AIBL test set. Participants are counted
uniquely within each interval; age is mean $\pm$ SD.}
\label{tab:dataset}

\setlength{\tabcolsep}{7pt}
\renewcommand{\arraystretch}{0.95}
\footnotesize

\begin{tabular}{l r c c}
\toprule
\textbf{Interval}
& \textbf{Pairs}
& \textbf{Participants}
& \textbf{Age (years)} \\
\midrule

\multicolumn{4}{l}{\textbf{ADNI (Internal Dataset)}} \\

2 y  & 3440 & 1434 & $73.3 \pm 7.0$ \\
3 y  & 1942 & 858  & $73.5 \pm 6.8$ \\
4 y  & 1568 & 676  & $73.0 \pm 6.6$ \\
5 y  & 1044 & 448  & $73.0 \pm 6.4$ \\
6 y  & 871  & 402  & $72.9 \pm 6.5$ \\
7 y  & 608  & 320  & $72.5 \pm 6.3$ \\
8 y  & 398  & 251  & $72.3 \pm 6.5$ \\
9 y  & 244  & 165  & $72.7 \pm 6.1$ \\
10 y & 149  & 103  & $72.7 \pm 6.3$ \\

\textbf{Total}
& \textbf{10264}
& \textbf{1468}
& $\mathbf{73.2 \pm 7.0}$ \\

\midrule

\multicolumn{4}{l}{\textbf{AIBL (External Test Set)}} \\

3 y & 275 & 147 & $72.6 \pm 6.9$ \\
6 y & 57  & 57  & $73.3 \pm 6.4$ \\

\textbf{Total}
& \textbf{332}
& \textbf{150}
& $\mathbf{72.6 \pm 6.8}$ \\

\bottomrule
\end{tabular}
\end{table}

\section{Method}

\subsection{One-Step Diffusion Model}

As illustrated in Figure~\ref{fig:method}, we propose a
covariate-conditioned one-step diffusion framework for
multi-horizon prediction of 3D GMD maps. Given baseline
and follow-up maps $x_b$ and $x_f$, the target change
is $\Delta x=x_f-x_b$. The model predicts
$\widehat{\Delta x}$, and the future map is reconstructed
as $\hat{x}_f=x_b+\widehat{\Delta x}$.
This residual formulation exploits anatomical
correspondence between visits, focusing prediction
on longitudinal changes rather than complete anatomy.

We adopt a denoising diffusion probabilistic model
(DDPM)~\cite{ho2020denoising} with $T=1000$ timesteps
and a linear noise schedule. At a uniformly sampled
training timestep $t$, the ground-truth change is
perturbed according to
\begin{equation}
x_t=\sqrt{\bar{\alpha}_t}\,\Delta x+
\sqrt{1-\bar{\alpha}_t}\,\epsilon,
\quad \epsilon\sim\mathcal{N}(0,I),
\label{eq:forward}
\end{equation}
where $x_t$ is the noisy change, $\epsilon$ is
standard Gaussian noise, $I$ is the identity matrix,
and $\bar{\alpha}_t=\prod_{s=0}^{t}(1-\beta_s)$,
with $\beta_s$ denoting the noise variance.
Using $x_0$-parameterization, the network directly
estimates the clean change:
$\widehat{\Delta x}_t=f_\theta(x_t,x_b,c,t)$,
where $f_\theta$ is the denoising network,
$\theta$ its parameters, and $c$ the covariate vector.

Our network is a 3D adaptation of the ADM
U-Net~\cite{dhariwal2021adm}, with 128 base channels,
multipliers $(1,2,4,8)$, two residual blocks per
resolution level, and self-attention at the deepest
resolution, totaling approximately 716M parameters.

Conditioning combines anatomical and participant-specific
information. Anatomical conditioning concatenates the
clean baseline GMD map, rescaled to $[-1,1]$, with
the noisy change $x_t$ as separate input channels. A six-dimensional covariate vector $c$ encodes
normalized baseline age, one-hot diagnosis
(CN, MCI, dementia), APOE4 allele count
(0, 1, or 2), and continuous follow-up interval. 
A multilayer perceptron (MLP) projects the
six-dimensional covariate vector $c$ into the
512-dimensional timestep embedding space,
where it is added to condition the U-Net
residual blocks.

At inference, Gaussian noise
$x_{999}\sim\mathcal{N}(0,I)$ is supplied with the
baseline map and covariates. A single evaluation,
$\widehat{\Delta x}=f_\theta(x_{999},x_b,c,999)$,
predicts the change, which is added to $x_b$ to
reconstruct the future GMD map without iterative
reverse diffusion sampling.

\subsection{Training Objective}

The training objective combines voxel-wise mean
squared error (MSE) with a spatial correlation
penalty. Let $\Omega_{\mathrm{GM}}$ denote the
gray-matter mask and $|\Omega_{\mathrm{GM}}|$
its voxel count. The masked MSE is defined as
\begin{equation}
\mathcal{L}_{\mathrm{MSE}}=
\frac{1}{|\Omega_{\mathrm{GM}}|}
\sum_{i\in\Omega_{\mathrm{GM}}}
(\widehat{\Delta x}_{t,i}-\Delta x_i)^2,
\label{eq:mse}
\end{equation}
where $\widehat{\Delta x}_{t,i}$ and $\Delta x_i$
represent the predicted and true changes at voxel $i$.
The spatial Pearson correlation is defined as
\begin{equation}
r_{\mathrm{GM}} =
\frac{
\sum_{i\in\Omega_{\mathrm{GM}}}
(\widehat{\Delta x}_{t,i}-\overline{\widehat{\Delta x}_t})
(\Delta x_i-\overline{\Delta x})
}{
\sqrt{
\sum_{i\in\Omega_{\mathrm{GM}}}
(\widehat{\Delta x}_{t,i}-\overline{\widehat{\Delta x}_t})^2
\sum_{i\in\Omega_{\mathrm{GM}}}
(\Delta x_i-\overline{\Delta x})^2
}},
\label{eq:pearson}
\end{equation}
where overbars denote means within the gray-matter
mask. The corresponding correlation penalty is
$\mathcal{L}_{\mathrm{corr}}=1-r_{\mathrm{GM}}$.
The total loss is
\begin{equation}
\mathcal{L}=
\mathcal{L}_{\mathrm{MSE}}+
0.1\,\mathbb{I}(t>500)\mathcal{L}_{\mathrm{corr}},
\label{eq:loss}
\end{equation}
with $\mathbb{I}(t>500)$ activating the correlation
penalty at higher noise levels. Both loss terms
are computed within the gray-matter mask, which
is not provided as a network input.

\subsection{Experimental Setup}

The model was trained on ADNI using the configuration
in Table~\ref{tab:experimental_setup}, with
participant-level splitting to prevent data leakage.
Hyperparameters were selected on the validation set,
and the model was evaluated on held-out ADNI and
external AIBL data without retraining.

Following~\cite{farki2026forecasting}, we use
$\Delta$-Pearson to measure spatial agreement
between predicted and true longitudinal changes.
It is computed on the final predictions using
the Pearson correlation defined in
Equation~\eqref{eq:pearson}. Mean absolute error
(MAE) and MSE measure voxel-wise change errors,
while reconstructed-image Pearson correlation,
peak signal-to-noise ratio (PSNR), and structural
similarity index (SSIM) assess image fidelity.
All metrics were computed per test pair within
the gray-matter mask and averaged across pairs.
Regional $\Delta$-Pearson was evaluated in the
hippocampus and entorhinal cortex, two regions
particularly relevant to neurodegeneration,
using anatomical regions defined by the
Neuromorphometrics atlas.

\begin{table}[t]
\centering
\caption{Summary of the experimental configuration.}
\label{tab:experimental_setup}

\setlength{\tabcolsep}{3pt}
\renewcommand{\arraystretch}{0.92}
\footnotesize

\begin{tabular}{p{0.23\columnwidth} p{0.69\columnwidth}}
\toprule
\textbf{Setting} & \textbf{Configuration} \\
\midrule

Data split
& Participant-level: 1,028 train / 220 validation /
220 test; nine follow-up intervals \\

Optimiser
& AdamW, learning rate $1.4\times10^{-4}$ linearly
annealed to 0, gradient clipping 1.0 \\

Training
& Batch size 4, FP16, EMA 0.9999, 30,000 steps \\

Augmentation
& Random left--right flip (training only) \\

Hardware
& $2\times$ NVIDIA GH200, $\sim$50 h training \\

\bottomrule
\end{tabular}
\end{table}

\subsection{Experimental Results}

Table~\ref{tab:main_results} summarizes performance across
prediction horizons. On ADNI, the model achieved an overall
$\Delta$-Pearson of 0.478, MSE of $5.176\times10^{-3}$,
and SSIM of 0.979. Interestingly, $\Delta$-Pearson increased
from 0.444 at 2 years to 0.582 at 10 years, whereas MSE
and MAE generally increased with horizon. This may reflect
more pronounced spatial patterns of change at longer
intervals, despite greater errors in their predicted
magnitudes. Regional $\Delta$-Pearson reached 0.503
in the hippocampus and 0.509 in the entorhinal cortex,
indicating spatial agreement in regions relevant to
neurodegeneration. External evaluation on AIBL yielded
$\Delta$-Pearson of 0.450, MSE of $4.016\times10^{-3}$,
and SSIM of 0.985, demonstrating transfer to an
independent cohort without retraining. One-step
inference required 1.451\,s per 3D volume.

\begin{figure*}[!b]
    \centering
    \includegraphics[width=0.86\textwidth]{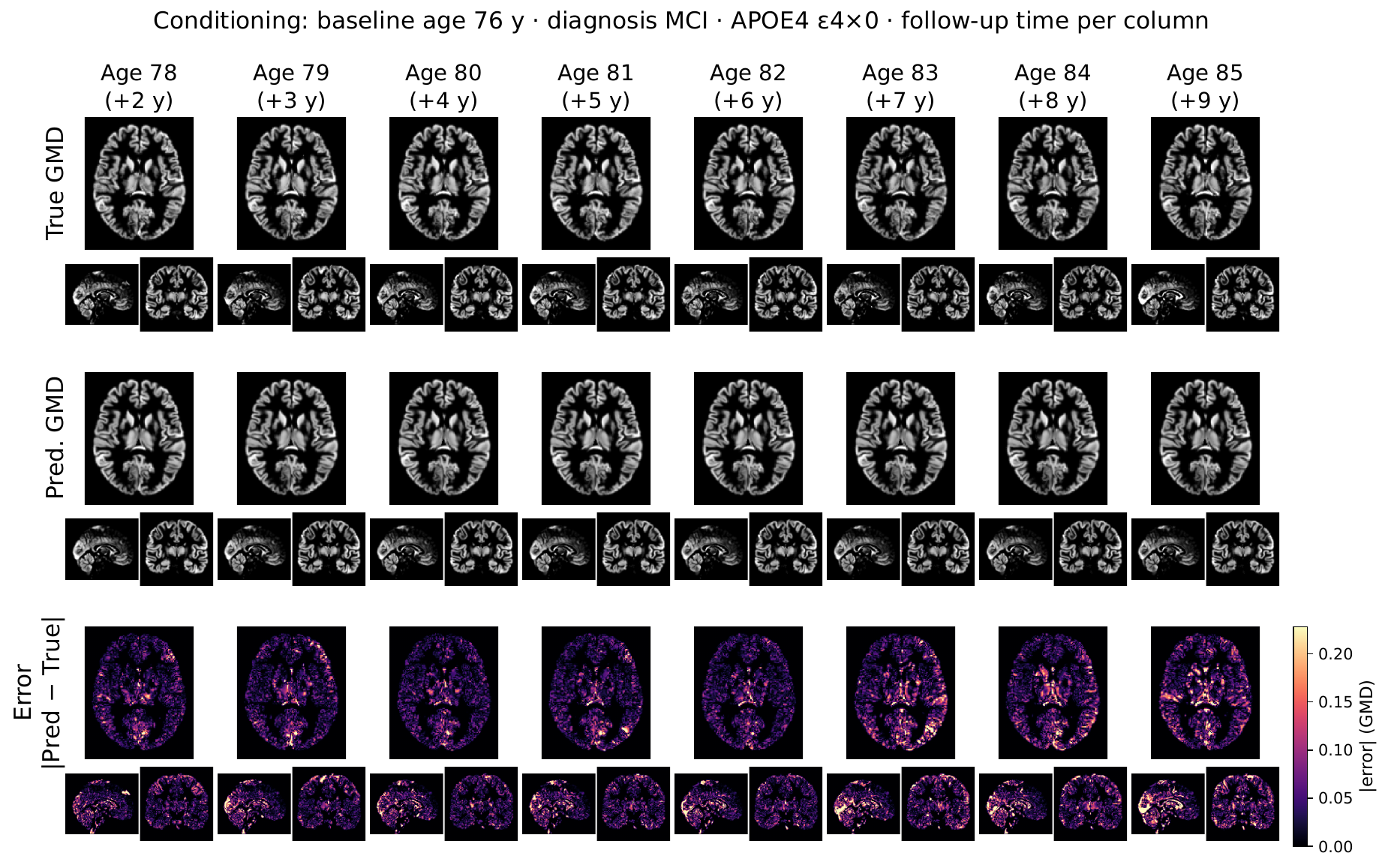}
    \caption{
Qualitative multi-horizon GMD predictions.
Columns represent follow-up intervals from 2 to 9 years,
while rows show ground-truth GMD maps, predicted maps,
and absolute error maps in axial, sagittal, and coronal
views. }

    \label{fig:qualitative}
\end{figure*}

\begin{table*}[!t]
\centering
\footnotesize
\setlength{\tabcolsep}{2.2pt}
\renewcommand{\arraystretch}{0.92}

\caption{Multi-horizon performance on the ADNI main test set and the external AIBL test set. whole-brain change metrics are reported as mean $\pm$ SD.}
\label{tab:main_results}

\begin{tabular}{c ccc ccc cc}
\toprule
& \multicolumn{3}{c}{\textbf{Whole-Brain Change}}
& \multicolumn{3}{c}{\textbf{Reconstructed GMD}}
& \multicolumn{2}{c}{\textbf{Regional $\Delta$-Pearson}} \\
\cmidrule(lr){2-4}
\cmidrule(lr){5-7}
\cmidrule(lr){8-9}

\textbf{Horizon}
& $\Delta$-Pearson $\uparrow$
& \shortstack{MSE $\downarrow$\\($\times 10^{-3}$)}
& \shortstack{MAE $\downarrow$\\($\times 10^{-2}$)}
& Pearson $\uparrow$
& \shortstack{PSNR $\uparrow$\\(dB)}
& SSIM $\uparrow$
& \shortstack{{Hippocampus}\\$\uparrow$}
& \shortstack{{Entorhinal}\\{Cortex} $\uparrow$} \\

\midrule
\multicolumn{9}{l}{\textbf{ADNI (Main Test Set)}} \\

2 y
& $0.444 \pm 0.060$ & $4.322 \pm 1.954$ & $4.622 \pm 0.987$
& 0.936 & 26.124 & 0.982 & 0.455 & 0.479 \\

3 y
& $0.464 \pm 0.067$ & $4.959 \pm 2.340$ & $4.951 \pm 1.118$
& 0.927 & 25.543 & 0.980 & 0.494 & 0.503 \\

4 y
& $0.478 \pm 0.068$ & $5.313 \pm 2.579$ & $5.118 \pm 1.194$
& 0.923 & 25.367 & 0.978 & 0.513 & 0.515 \\

5 y
& $0.498 \pm 0.072$ & $5.916 \pm 2.900$ & $5.435 \pm 1.285$
& 0.914 & 24.972 & 0.976 & 0.531 & 0.536 \\

6 y
& $0.512 \pm 0.070$ & $6.007 \pm 2.818$ & $5.469 \pm 1.226$
& 0.913 & 24.914 & 0.976 & 0.552 & 0.534 \\

7 y
& $0.527 \pm 0.067$ & $6.309 \pm 2.962$ & $5.618 \pm 1.261$
& 0.909 & 24.832 & 0.974 & 0.564 & 0.545 \\

8 y
& $0.546 \pm 0.067$ & $6.276 \pm 3.130$ & $5.586 \pm 1.295$
& 0.910 & 24.964 & 0.975 & 0.577 & 0.544 \\

9 y
& $0.562 \pm 0.069$ & $6.675 \pm 3.158$ & $5.789 \pm 1.270$
& 0.904 & 24.529 & 0.973 & 0.584 & 0.560 \\

10 y
& $0.582 \pm 0.059$ & $6.041 \pm 2.219$ & $5.502 \pm 0.962$
& 0.909 & 24.819 & 0.975 & 0.576 & 0.527 \\

\textbf{Overall}
& $\mathbf{0.478 \pm 0.074}$
& $\mathbf{5.176 \pm 2.574}$
& $\mathbf{5.053 \pm 1.195}$
& \textbf{0.924}
& \textbf{25.498}
& \textbf{0.979}
& \textbf{0.503}
& \textbf{0.509} \\

\midrule
\multicolumn{9}{l}{\textbf{AIBL (External Test Set)}} \\

3 y
& $0.444 \pm 0.057$ & $3.951 \pm 2.096$ & $4.369 \pm 0.983$
& 0.948 & 26.364 & 0.985 & 0.461 & 0.470 \\

6 y
& $0.474 \pm 0.066$ & $4.327 \pm 2.434$ & $4.578 \pm 0.937$
& 0.943 & 26.121 & 0.984 & 0.513 & 0.498 \\

\textbf{Overall}
& $\mathbf{0.450 \pm 0.060}$
& $\mathbf{4.016 \pm 2.163}$
& $\mathbf{4.405 \pm 0.978}$
& \textbf{0.947}
& \textbf{26.322}
& \textbf{0.985}
& \textbf{0.470}
& \textbf{0.475} \\

\bottomrule
\end{tabular}

\vspace{7pt}

\caption{Ablation study on the ADNI test set.  whole-brain change metrics are reported as mean $\pm$ SD.}
\label{tab:ablation}

\begin{tabular}{l ccc ccc cc}
\toprule
& \multicolumn{3}{c}{\textbf{Whole-Brain Change}}
& \multicolumn{3}{c}{\textbf{Reconstructed GMD}}
& \multicolumn{2}{c}{\textbf{Regional $\Delta$-Pearson}} \\
\cmidrule(lr){2-4}
\cmidrule(lr){5-7}
\cmidrule(lr){8-9}

\textbf{Configuration}
& $\Delta$-Pearson $\uparrow$
& \shortstack{MSE $\downarrow$\\($\times 10^{-3}$)}
& \shortstack{MAE $\downarrow$\\($\times 10^{-2}$)}
& Pearson $\uparrow$
& \shortstack{PSNR $\uparrow$\\(dB)}
& SSIM $\uparrow$
& \shortstack{{Hippocampus}\\$\uparrow$}
& \shortstack{{Entorhinal}\\{Cortex} $\uparrow$} \\

\midrule

\textbf{Full model}
& $\mathbf{0.478 \pm 0.074}$
& $\mathbf{5.176 \pm 2.574}$
& $\mathbf{5.053 \pm 1.195}$
& \textbf{0.924}
& \textbf{25.498}
& \textbf{0.979}
& \textbf{0.503}
& \textbf{0.509} \\

w/o Corr. loss
& $0.456 \pm 0.070$
& $5.341 \pm 2.665$
& $5.128 \pm 1.233$
& 0.922 & 25.369 & 0.978 & 0.477 & 0.493 \\

w/o covariates
& $0.461 \pm 0.070$
& $5.399 \pm 2.795$
& $5.138 \pm 1.286$
& 0.922 & 25.349 & 0.978 & 0.477 & 0.490 \\

Image pred.
& $0.452 \pm 0.074$
& $5.408 \pm 2.592$
& $5.266 \pm 1.202$
& 0.923 & 25.278 & 0.978 & 0.464 & 0.488 \\

\bottomrule
\end{tabular}

\end{table*}

Figure~\ref{fig:qualitative} illustrates our model's predictions for an ADNI test participant selected for having the largest number of distinct follow-up intervals (eight) from a single baseline scan, showing preserved anatomy and larger errors at longer horizons, consistent with increasing MAE in Table~\ref{tab:main_results}.

Table~\ref{tab:ablation} presents an ablation study
evaluating the contributions of the correlation loss,
covariate conditioning, and residual prediction. Removing the correlation loss reduced
$\Delta$-Pearson from 0.478 to 0.456, suggesting that
voxel-wise MSE alone does not fully capture the spatial
organization of longitudinal changes. Removing covariates reduced $\Delta$-Pearson to 0.461,
demonstrating the value of additional subject-specific information.
Direct image prediction yielded the lowest $\Delta$-Pearson
(0.452), highlighting the benefit of explicitly modeling longitudinal differences.

\section{Conclusion}

We presented a covariate-conditioned one-step diffusion
framework for multi-horizon prediction of longitudinal
3D GMD changes. The model captures spatial patterns
of brain changes and generalizes from ADNI to AIBL
without retraining. Ablation results support the benefits
of covariate conditioning, correlation-based supervision,
and residual prediction. However, uneven representation
across follow-up intervals and diagnostic groups limits
evaluation of underrepresented populations. Future work
will explore larger and more balanced cohorts to improve
longitudinal prediction.

\FloatBarrier
\clearpage
\section*{Compliance with Ethical Standards}

This retrospective study used data from ADNI \cite{adni_data} and AIBL \cite{aibl_data}, which received
ethical approval and participant informed consent. Secondary analysis
required no additional ethical approval under the respective data-use
agreements.

\section*{Acknowledgments}

Data were obtained from ADNI
(\nolinkurl{https://adni.loni.usc.edu/}) and AIBL
(\nolinkurl{https://data.aibl.org.au/adni/}).
Computational resources were provided by
CSC -- IT Center for Science, Finland,
and the UEF Bioinformatics Center,
Biocenter Kuopio, and Biocenter Finland
at the University of Eastern Finland.
This work was supported by the Research
Council of Finland (grants 346934,
PRIMAL, and 358944, Flagship of Advanced
Mathematics for Sensing, Imaging and
Modeling), with additional support from
the European Union through the Erasmus
Mundus Joint Master programme in
Intelligent Photonics for Security,
Reliability, Sustainability and Safety
(iPSRS). OpenAI's GPT-5 was used for
language editing and readability; all
content was reviewed and approved by
the authors.

\bibliographystyle{IEEEtran}
\bibliography{strings,refs}

@inproceedings{farki2026forecasting,
  author    = {Farki, Ali and Moradi, Elaheh and Koundal, Deepika and Tohka, Jussi},
  title     = {Forecasting Future Anatomies: Longitudinal Brain {MRI}-to-{MRI} Prediction},
  booktitle = {2026 IEEE 23rd International Symposium on Biomedical Imaging (ISBI)},
  pages     = {1--5},
  year      = {2026},
  publisher = {IEEE}
}

@inproceedings{litrico2024tadm,
  author    = {Litrico, Mattia and Guarnera, Francesco and Giuffrida, Mario Valerio and Rav{\`i}, Daniele and Battiato, Sebastiano},
  title     = {{TADM}: Temporally-Aware Diffusion Model for Neurodegenerative Progression on Brain {MRI}},
  booktitle = {Medical Image Computing and Computer-Assisted Intervention (MICCAI)},
  pages     = {444--453},
  year      = {2024},
  publisher = {Springer}
}

@article{puglisi2025brain,
  author  = {Puglisi, Lemuel and Alexander, Daniel C. and Rav{\`i}, Daniele},
  title   = {Brain Latent Progression: Individual-Based Spatiotemporal Disease Progression on 3D Brain {MRIs} via Latent Diffusion},
  journal = {Medical Image Analysis},
  pages   = {103734},
  year    = {2025}
}

@article{hedman2012human,
  author  = {Hedman, Anna M. and van Haren, Neeltje E. M. and Schnack, Hugo G. and Kahn, Ren{\'e} S. and Hulshoff Pol, Hilleke E.},
  title   = {Human Brain Changes Across the Life Span: A Review of 56 Longitudinal Magnetic Resonance Imaging Studies},
  journal = {Human Brain Mapping},
  volume  = {33},
  number  = {8},
  pages   = {1987--2002},
  year    = {2012}
}

@article{song2020denoising,
  author  = {Song, Jiaming and Meng, Chenlin and Ermon, Stefano},
  title   = {Denoising Diffusion Implicit Models},
  journal = {arXiv preprint arXiv:2010.02502},
  year    = {2020}
}

@inproceedings{he2025lotus,
  author    = {He, Jing and Li, Haodong and Yin, Wei and Liang, Yixun and Li, Leheng and Zhou, Kaiqiang and Zhang, Hongbo and Liu, Bingbing and Chen, YingCong},
  title     = {Lotus: Diffusion-Based Visual Foundation Model for High-Quality Dense Prediction},
  booktitle = {International Conference on Learning Representations},
  year      = {2025}
}

@article{ye2024stablenormal,
  author  = {Ye, Chongjie and Qiu, Lingteng and Gu, Xiaodong and Zuo, Qi and Wu, Yushuang and Dong, Zilong and Bo, Liefeng and Xiu, Yuliang and Han, Xiaoguang},
  title   = {{StableNormal}: Reducing Diffusion Variance for Stable and Sharp Normal},
  journal = {ACM Transactions on Graphics},
  volume  = {43},
  number  = {6},
  pages   = {1--18},
  year    = {2024}
}

@article{wu2025stableintrinsic,
  author  = {Wu, Xiuchao and Zhu, Pengfei and Lyu, Jiangjing and Liu, Xinguo and Guo, Jie and Guo, Yanwen and Xu, Weiwei and Lyu, Chengfei},
  title   = {{StableIntrinsic}: Detail-Preserving One-Step Diffusion Model for Multi-View Material Estimation},
  journal = {arXiv preprint arXiv:2508.19789},
  year    = {2025}
}

@inproceedings{dhariwal2021adm,
  author    = {Dhariwal, Prafulla and Nichol, Alex},
  title     = {Diffusion Models Beat {GANs} on Image Synthesis},
  booktitle = {Advances in Neural Information Processing Systems},
  volume    = {34},
  pages     = {8780--8794},
  year      = {2021}
}

@article{lorenzi2015disentangling,
  author  = {Lorenzi, Marco and Pennec, Xavier and Frisoni, Giovanni B. and Ayache, Nicholas},
  title   = {Disentangling Normal Aging from Alzheimer's Disease in Structural Magnetic Resonance Images},
  journal = {Neurobiology of Aging},
  volume  = {36},
  pages   = {S42--S52},
  year    = {2015}
}

@article{tondelli2012structural,
  author  = {Tondelli, Manuela and Wilcock, Gordon K. and Nichelli, Paolo and De Jager, Celeste A. and Jenkinson, Mark and Zamboni, Giovanna},
  title   = {Structural {MRI} Changes Detectable up to Ten Years Before Clinical Alzheimer's Disease},
  journal = {Neurobiology of Aging},
  volume  = {33},
  number  = {4},
  pages   = {825.e25},
  year    = {2012}
}

@article{hadji2026predicting,
  author  = {Hadji, M. and Moradi, E. and Tohka, J.},
  title   = {Predicting Future Brain Atrophy Based on Longitudinal {MRI}},
  journal = {Journal of Alzheimer's Disease},
  volume  = {113},
  number  = {2},
  pages   = {796--812},
  year    = {2026},
  doi     = {10.1177/13872877261471049}
}

@article{ho2020denoising,
  author  = {Ho, Jonathan and Jain, Ajay and Abbeel, Pieter},
  title   = {Denoising Diffusion Probabilistic Models},
  journal = {Advances in Neural Information Processing Systems},
  volume  = {33},
  pages   = {6840--6851},
  year    = {2020}
}

@article{gaser2024cat,
  author  = {Gaser, C. and Dahnke, R. and Thompson, P. M. and Kurth, F. and Luders, E.},
  title   = {{CAT}: A Computational Anatomy Toolbox for the Analysis of Structural {MRI} Data},
  journal = {GigaScience},
  volume  = {13},
  pages   = {giae049},
  year    = {2024},
  doi     = {10.1093/gigascience/giae049}
}

@misc{cat12website,
  author       = {{Structural Brain Mapping Group}},
  title        = {{CAT12}: Computational Anatomy Toolbox},
  howpublished = {\url{https://neuro-jena.github.io/cat/}},
  note         = {Jena University Hospital. Accessed: October 2026}
}

@misc{adni_data,
  author = {{Alzheimer's Disease Neuroimaging Initiative}},
  title = {{ADNI Data}},
  url = {https://adni.loni.usc.edu/data-samples/adni-data/}
}

@misc{aibl_data,
  author = {{Australian Imaging, Biomarkers and Lifestyle Study}},
  title = {{Collaboration and Accessing AIBL Data}},
  url = {https://data.aibl.org.au/adni/}
}

\end{document}